\documentclass[12pt,preprint]{aastex}
\everymath{\rm}
\received{}
\revised{}
\accepted{}
\cpright{}{}\journalid{}{}
\articleid{}{}
\paperid{}
\slugcomment{To appear in {\it The Astrophysical Journal Supplement}}
\shortauthors{Kwitter \& Henry}
\shorttitle{S, Cl, \& Ar in PNe}
\begin{document}
\title{Sulfur, Chlorine, \& Argon Abundances 
in Planetary Nebulae. IIB: Abundances in a
Southern Sample}

\author{J.B. Milingo\footnote{Visiting Astronomer, Cerro Tololo Interamerican 
Observatory, National Optical Astronomy Observatories,
which is operated by the Association of Universities for Research in
Astronomy, Inc. (AURA) under cooperative agreement with the
National Science Foundation.}}

\affil{Department of Physics, Gettysburg College, Gettysburg, PA
17325; jmilingo@gettysburg.edu}

\author {R.B.C. Henry$^1$}

\affil{Department of Physics \& Astronomy, University of Oklahoma,
Norman, OK 73019; henry@mail.nhn.ou.edu}

\and

\author{K.B. Kwitter}

\affil{Department of Astronomy, Williams College, Williamstown, MA
01267; kkwitter@williams.edu}

\begin{abstract}
We have undertaken a large spectroscopic survey of over 80
planetary nebulae with the goal of providing a homogeneous
spectroscopic database between 3600-9600~{\AA}, as well as a set of
consistently determined abundances, especially for oxygen, sulfur,
chlorine, and argon.  In the current paper we calculate and report the
S/O, Cl/O, and Ar/O abundance ratios for 45 southern planetary nebulae
(predominantly Type~II), using our own recently observed line
strengths published in an earlier paper. One of the salient features
of our work is the use of the NIR lines of [S~III]
$\lambda\lambda$9069,9532 coupled with the [S~III] temperature, to
determine the S$^{+2}$ ionic abundance. We find the following average abundances
for these objects: S/O=0.011$\pm$.0064, Cl/O=0.00031$\pm$.00012, and
Ar/O=0.0051$\pm$.0020.

\end{abstract}


\clearpage

\section{Introduction}

The abundances of S, Cl, and Ar relative to O are useful in that they
provide valuable constraints on nucleosynthesis predictions of
theoretical stellar evolution models used in the study of the chemical
evolution of galaxies. We are currently carrying out a 
spectrophotometric survey of over 80 (mostly Type~II)
planetary nebulae with the hope of producing a homogeneous database 
extending 
from 3600-9600~{\AA}. Our goal is to calculate abundances of sulfur, chlorine, 
and argon in a consistent fashion in an effort to compare our empirical results 
with theoretical yield predictions for these elements.

In the first paper in this series, Kwitter \&
Henry (2001), hereafter Paper~I, we presented new spectrophotometric
measurements extending from 3600-9600~{\AA} as well as abundance
calculations for 19 planetary nebulae (PNe), 14 Type~II and five that
our results either confirmed or newly indicated may be Type~I. This broad
spectral coverage allowed us to observe the strong nebular lines of
[S~III] $\lambda\lambda$9069,9532 for a large sample of objects and to
use these lines to compute abundances of S$^{+2}$.  Also in that
paper, we carefully evaluated the atomic data as well as the ionization
correction factors (ICF) relevant to abundances of S, Cl, and Ar, and
developed a new ICF for sulfur using a grid of detailed
photoionization models. The reader is directed to Paper~I for the
details pertaining to the motivation and particulars of our entire study.

The current paper is a continuation of this project. We use
spectrophotometric observations published in Milingo et al. (2001;
hereafter MKHC) of 45 southern PNe, including 38 Type~II (three in
common with Paper~I) and seven that our results either confirm or newly
indicate may be Type~I (four in common with Paper~I), to calculate ionic
and elemental abundances for several elements including O, S, Cl, and
Ar.

Section 2 contains a brief review of the methods used to obtain the
abundances, while our results are given in {\S}3. Section~4 is a
summary of our findings.

\section{Calculations}

Electron temperatures and densities, ionic abundances, and total
elemental abundances were computed from our measured line strengths
using the procedure developed and employed in Paper~I, to which
the reader is referred for a more extended discussion. Briefly, the
method makes use of the program ABUN (written by R.B.C.H. and described in
\S~3.1 of Paper~I), which
features a 5-level atom routine along with ionization correction
factors (ICF) outlined below. Table~1 summarizes the ions
observed, the wavelengths of the emission lines used to obtain ionic
abundances, temperatures, and densities, and the sources of the atomic
data used in ABUN.

As in Paper~I, the present study is distinguished by its extended use
of the NIR lines of [S~III] $\lambda\lambda$9069,9532 for determining
S$^{+2}$ abundances. These nebular lines are much stronger than the
auroral line at $\lambda$6312, yet ostensibly suffer from effects of
telluric absorption and emission, effects that are difficult
to remove completely. Since the intensity ratio of $\lambda$9532/$\lambda$9069 is
determined by atomic physics to be 2.48 (Mendoza 1983), we simply assumed that if the
observed ratio for an object was greater than or equal to 2.48, then
the $\lambda$9532 line was less affected by atmospheric absorption,
and we used it in the ionic abundance calculation for
S$^{+2}$. Alternatively, when the observed line ratio was less than
2.48, we used the $\lambda$9069 line for the ionic abundance
determination, assuming it to be the less affected. More discussion is
given on this point in Paper~I.

Input line strengths for the program ABUN were taken from Tables~2A-H
in MKHC. Electron temperatures and densities were calculated in
the standard fashion using ratios of lines from auroral and nebular
transitions sensitive to temperature, density, or both. Abundances of
observed ions were calculated by dividing the energy production rate
per ion at the observed wavelength $\epsilon_{\lambda}$(T$_e$,N$_e$)
into the observed line intensity I$_{\lambda}$ for each of the
emission lines listed in Table~1. In calculating ionic abundances, the
[O~III] temperature was used for the high ionization species
(O$^{+2}$, Ne$^{+2}$, Cl$^{+2}$, Cl$^{+3}$, Ar$^{+2}$, and Ar$^{+3}$),
the [N~II] temperature was used for the low ionization species (O$^o$,
O$^+$, N$^+$, and S$^+$), and the [S~III] temperature was used for
S$^{+2}$. For all species, we employed the [S~II] density, the only density
that we were able to derive from our data.

Final elemental abundances were calculated by adding the abundances of
observed ions for an element and then multiplying this sum by the
appropriate ionization correction factor. An extended discussion of
ICFs is presented in Paper~I and will not be repeated here.

\section{Results}

Ionic abundances, ICFs, and electron temperatures and densities
resulting from the above calculative procedure are given in
Tables~2A-H. Note that together the tables report on 45 individual
PNe, with one entry each for the two positions observed for NGC~3242
in Table~2F. A guide to the uncertainties of these tabulated results
is given in the footnotes of each table. In several instances, changes
in the procedure for calculating abundances had to be made in order to
accommodate missing line strengths or unrealistic electron
temperatures. First, in cases where [O~III]
$\lambda$4363 was not observed, we employed a value for T$_{O3}$
estimated from T$_{N2}$, using a relation derived from the grids of
photoionization models discussed in Paper~I, i.e.,
\begin{equation}
T_{O3}=7414-1.058T_{N2}+0.0001322 T_{N2}^2
\end{equation}
Such cases are indicated with footnotes in Tables~2A-H. Second, in a
few instances derived electron temperatures seemed unrealistic and we
indicate these cases by reporting the temperature inside a set of
parentheses. When this situation occurred with T$_{S3}$, T$_{O3}$ was
used instead to obtain the values for S$^{+2}$/H$^+$$_{NIR}$. These
cases are also indicated with footnotes.  Finally, when $\log (O/O^+)\le +0.6$, 
then
Ar=Ar$^{+2}$, and we entered an ellipsis
for the relevant object on the ICF(Ar) line in Tables~2A-H.

Since sulfur is a major focus of this set of papers, we compare the
results for S$^{+2}$ as determined from the [S~III] $\lambda$6312 line
and from the NIR lines of [S~III] in Fig.~1. Clearly there is a tendency for the S$^{+2}$
abundances derived from the 6312~{\AA} line to be greater, in many
cases by a factor of two. This trend is better defined now than it was
in Paper~I, where we could only plot 19 objects. While we believe that this
discrepancy may be related to our use of T$_{S3}$ to calculate
S$^{+2}$ abundances (since T$_{S3}$ is often higher than T$_{N2}$ or
T$_{O3}$, and thus gives lower abundances), we will analyze this
situation in much more detail in a later paper when we have completed
our consideration of all of the objects in our entire sample.

Our elemental abundances are summarized in Tables~3A-H. The first
column in each table identifies the abundance ratio being tabulated,
while subsequent columns give our results for the individual PNe
listed across the top. Note that the last two columns give published
values for the Sun from Grevesse et al. (1996) and the Orion Nebula
from Esteban et al. (1998). Uncertainties are indicated in the
tablenotes. Those cases mentioned above where the abundance of Ar was
set equal to the abundance of Ar$^{+2}$ are also indicated with
footnotes.

Using the discriminant given by Kingsburgh \& Barlow (1994), in which
a Galactic Type~I PN is defined to have N/O$>$0.8, we find that seven
of our objects meet this criterion: He2-123, IC~2621, PB6, M1-54,
Pe1-18, NGC~6439 and M1-57. The last four objects are also included in
Paper~I; M1-54 and Pe1-18 have N/O ratios above 0.8 in both analyses;
NGC~6439 and M1-57 have N/O greater than 0.8 in either Paper~I or this
paper (0.90 vs. 0.71 and 1.35 vs. 0.69, respectively.  Since the
values in each case that are below 0.8 are not below by much, we
classify all four of these PNe as probable Type~I. (Note that the
values quoted for the northern observations in Table~5 below for
Pe1-18 and M1-57 are correct and supersede values in Paper~I.)

Fig.~2 provides plots of S/O, Cl/O and Ar/O versus O/H for the objects
in this paper (filled circles) and Paper~I (open circles).  We note the 
larger scatter in the
cases of Cl/O and Ar/O, which is undoubtedly due to the weaker, more
uncertain line strengths associated with Cl and Ar than with
S. 

Table~4 lists the averages for several abundance ratios and
samples. Although our sample contains mostly Type~II PNe, there are in fact seven Type~Is in the group. In the second column, then, we report the averages for the entire sample, including the Type~Is, while in the third column we give the averages for the Type~I objects only. Type~I progenitors are assumed to be more massive than those of Type~II and thus come from a younger population with higher metallicity. Commensurate with this notion, the average values of O/H, S/H, Cl/H, and Ar/H (which should all track metallicity) for the seven Type~I objects are slightly higher than the results for the entire sample.

Next, note the excellent agreement between the current sample
and the one in Paper~I, which is encouraging given that the two sets of data were obtained on different telescopes (the 1.5~m telescope at CTIO for the current set and the KPNO 2.1~m telescope for the latter) and were reduced and measured independently. 

We again see the
tendency for S/H and S/O to run a bit lower than the averages obtained
by the AK sample as well as in the Sun and Orion, although we continue
to agree within the spread of points with the KB averages for
sulfur. Ours is the first major survey of S abundances which employs the NIR [S~III] lines along with [S~III] temperatures to infer S$^{+2}$ ion abundances. Earlier studies have used the [S~III] $\lambda$6312 line (when available) and either [N~II] or [O~III] temperatures to determine the abundance of this ion; in some cases the ion abundance could not be determined at all due to limitations in the data. We do note that generally our [S~III] temperatures are slightly greater than the temperatures used by others ([N~II] or [O~III]) when determining the abundance of S$^{+2}$ and this could be partially responsible for the offset between our results and others'.

Finally, Table~5 shows the ratios determined for the objects in
common with Paper~I in the form northern/southern. Note the generally excellent agreement for these seven objects.

\section{Summary}

We have used new spectrophotometric measurements published in Milingo
et al. (2001) to determine abundances of S, Cl, and Ar with
respect to both H and O for 45 planetary nebulae.  This
exercise is part of a larger project whose aim is to gather empirical
evidence for confronting theoretical predictions of nucleosynthesis by
detailed stellar evolution models. The distinguishing characteristic
of our work is that it uses the nebular [S~III] lines in the NIR at
$\lambda\lambda$9069,9532 to determine the S$^{+2}$ abundances. In
addition, in most cases, we are able to use the [S~III] temperature
(available now because we have observed both the auroral and nebular
lines) instead of the commonly used [O~III] or [N~II] temperatures. By
observing large samples of objects over the same broad wavelength
region and calculating ionic and element abundances using consistent
methods, we are working toward a goal of producing a large,
homogeneous set of observations and abundances for PNe.

We find values of S/O, Cl/O, and Ar/O in the present sample of 45
objects of 0.011$\pm$.0064, 0.00031$\pm$.00012, and 0.0051$\pm$.0020,
respectively. These agree remarkably well with the average values
obtained in Paper~I. At the same time, our calculated values of S/O
continue to be slightly lower systematically than those of previous
studies. This is probably due to the fact that most published S$^{+2}$
abundances, working from observations which usually do not go beyond
about 7000~{\AA} in the red, have been determined using the auroral
line of [S~III] $\lambda$6312 coupled with either the [O~III] or
[N~II] temperature, since the [S~III] temperature is unavailable. In the final
paper of this series we will address this problem in more detail,
using the complete sample of objects.

\acknowledgments

This research is supported by NSF grant AST-9819123.

\clearpage

\begin{deluxetable}{lccl}
\setlength{\tabcolsep}{0.04in}
\tabletypesize{\scriptsize}
\vspace{-2in} \tablecolumns{4} \tablewidth{0pc} \tablenum{1}
\tablecaption{ABUN: Sources Of Atomic Data} \tablehead{ Ion&Emission
Line ({\AA})&Data Type\tablenotemark{a}&Reference\tablenotemark{b} }
\startdata H$^0$ &4861& $\alpha_{eff}$($\lambda$4861) & 1 \\ He$^0$
&5876& $\alpha_{eff}$($\lambda$5876)\tablenotemark{c} & 2 \\ He$^+$
&4686& $\alpha_{eff}$($\lambda$4686) & 1\\ O$^o$ &6300& $\Omega$ & 3
\\ && A & 4 \\ O$^+$ &3727, 7323& $\Omega$ & 5 (2-3,4-5 only), 6 (all
other transitions) \\ && A & 4 \\ O$^{+2}$ &4363, 5007& $\Omega$ & 7
(4-5 only); 8 (all other transitions) \\ && A & 4 \\ N$^+$ &5755,
6584& $\Omega$ & 8 \\ && A & 4 \\ Ne$^{+2}$ &3869& $\Omega$ & 9 \\ &&
A & 10 \\ S$^{+}$ &4072, 6716, 6731& $\Omega$ & 11 \\ && A & 5 \\
S$^{+2}$ &6312, 9069, 9532& $\Omega$ & 12 \\ && A & 5 \\ Cl$^{+2}$
&5537&$\Omega$&13 \\ && A & 5 \\ Cl$^{+3}$ &8045&$\Omega$&12 \\ && A &
5 \\ Ar$^{+2}$ &7135& $\Omega$ & 12 \\ && A & 14 \\
Ar$^{+3}$&4740&$\Omega$&15 \\ && A & 16 \\ \enddata
\tablenotetext{a}{$\alpha_{eff}$=effective recombination coefficient;
$\Omega$=collision strength; A=transition rate.}
\tablenotetext{b}{References.--(1) Storey \& Hummer 1995; (2)
P{\'e}quignot et al.  1991; (3) Bhatia \& Kastner 1995; (4) Wiese,
Fuhr, \& Deters 1996; (5) Mendoza 1983; (6) McLaughlin \& Bell 1993;
(7) Burke, Lennon, \& Seaton 1989; (8) Lennon \& Burke 1994; (9)
Butler \& Zeippen 1994; (10) Baluja \& Zeippen 1988; (11) Ramsbottom,
Bell, \& Stafford 1996; (12) Galav{\'i}s et al. 1995; (13) Butler \&
Zeippen 1989; (14) Mendoza \& Zeippen 1983; (15) Zeippen, Le Bourlot,
\& Butler 1987; (16) Mendoza \& Zeippen 1982.}
\tablenotetext{c}{Includes collisional effects given by Clegg (1987).}
\end{deluxetable}

\begin{deluxetable}{lrrrrrr}
\tabletypesize{\scriptsize}
\setlength{\topmargin}{-1in}
\setlength{\tabcolsep}{0.03in}
\tablecolumns{7}
\tablewidth{0in}
\tablenum{2A}
\tablecaption{Ionic Abundances, Temperatures, \& Densities\tablenotemark{a}}
\tablehead{
\colhead{Parameter} & 
\colhead{Cn2-1} &
\colhead{Fg 1} &
\colhead{He2-21} &
\colhead{He2-37} &
\colhead{He2-48} &
\colhead{He2-55}  
}
\startdata

He$^+$/H$^+$&  0.13    &  0.11    &  9.05E-02&  6.20E-02&  9.75E-02&  3.25E-02 
\\
He$^{+2}$/H$^+$&  4.17E-03&  1.35E-02&  2.63E-02&  6.11E-02&  1.60E-02&  
9.46E-02 \\
ICF(He)&      1.00&      1.00&      1.00&      1.00&      1.00&      1.00 \\
O$^o$/H$^+$&  7.64E-06&  4.70E-06&  2.46E-06&  2.27E-05&  2.67E-05&    \nodata   
  \\
O$^+$/H$^+$&  2.04E-05&  3.31E-05&  1.18E-05&  1.34E-04&  1.07E-04&  8.07E-06 \\
O$^{+2}$/H$^+$&  6.71E-04&  3.01E-04&  2.20E-04&  4.19E-04&  2.56E-04&  1.68E-04 
\\
ICF(O)&      1.03&      1.12&      1.29&      1.99&      1.16&      3.91 \\
N$^{+}$/H$^+$&  7.95E-06&  1.42E-05&  1.73E-06&  4.30E-05&  3.34E-05&  1.64E-06 
\\
ICF(N)&     34.92&     11.27&     25.30&      8.18&      3.95&     85.12 \\
Ne$^{+2}$/H$^+$&  1.55E-04&  8.87E-05&  3.60E-05&  9.46E-05&  6.91E-05&  
4.34E-05 \\
ICF(Ne)&      1.06&      1.24&      1.36&      2.62&      1.65&      4.10 \\
S$^+$/H$^+$&  2.46E-07&  3.86E-07&  5.08E-08&  9.41E-07&  5.16E-07&  8.99E-08 \\
S$^{+2}$/H$^+_{NIR}$&  3.24E-06&  3.03E-06&  4.89E-07&  2.62E-06&  
1.01E-06\tablenotemark{b}&  1.26E-06 \\
S$^{+2}$/H$^+_{6312}$&  5.03E-06&  4.86E-06&  5.97E-06&  6.46E-06&    \nodata    
&  5.90E-06 \\
ICF(S)&      1.78&      1.34&      1.61&      1.28&      1.16&      2.63 \\
Cl$^{+2}$/H$^+$&  1.08E-07&  6.20E-08\tablenotemark{b}&  
2.06E-08\tablenotemark{b}&  9.27E-08&  7.08E-08\tablenotemark{c}&  
5.20E-08\tablenotemark{b} \\
Cl$^{+3}$/H$^+$&  6.43E-08&  4.88E-08\tablenotemark{b}&  4.40E-08&  5.72E-08&    
\nodata    &  4.07E-08\tablenotemark{c} \\
ICF(Cl)&      1.03&      1.12&      1.29&      1.99&      1.16&      3.91 \\
Ar$^{+2}$/H$^+$&  1.95E-06&  1.31E-06&  3.71E-07&  1.90E-06&  1.07E-06&  
7.78E-07 \\
Ar$^{+3}$/H$^+$&  6.07E-07&  4.27E-07&  4.37E-07&  2.01E-07&  
1.09E-07\tablenotemark{c}&  4.97E-07 \\
ICF(Ar)&      1.06&      1.23&      1.34&      2.26&      \nodata&      3.96 \\
T$_{O3}$(K)&     9600&     9300&    12200&    12000&    11200&    12200 \\
T$_{N2}$(K)&    10200&     8300&     7600&    10000&     9600&     9500 \\
T$_{O2}$(K)&    13700&     7100&    11500&    14500&        \nodata&    16100 \\
T$_{S2}$(K)&    13300&    (26100)&        \nodata&    10300&     6500&        
\nodata \\
T$_{S3}$(K)&    12000&     9400&    17600&    13800&        \nodata&    17200 \\
N$_{e,S2}$(cm$^{-3}$) &     4100&      600&     1500&      200&       10&      
200 \\

\enddata

\tablenotetext{a}{Unless otherwise noted, uncertainties in ionic abundances, 
electron temperatures, and electron densities are $\pm$30\%, $\pm$10\%, and 
$\pm$10\%, respectively.}

\tablenotetext{b}{Uncertainty is estimated to be $\pm$50\%}

\tablenotetext{c}{Uncertainty is estimated to be $\pm$75\%}

\end{deluxetable}

\begin{deluxetable}{lrrrrrr}
\tabletypesize{\scriptsize}
\setlength{\topmargin}{-1in}
\setlength{\tabcolsep}{0.03in}
\tablecolumns{7}
\tablewidth{0in}
\tablenum{2B}
\tablecaption{Ionic Abundances, Temperatures, \& Densities\tablenotemark{a}}
\tablehead{
\colhead{Parameter} & 
\colhead{He2-115} &
\colhead{He2-123} &
\colhead{He2-138} &
\colhead{He2-140} &
\colhead{He2-141} &
\colhead{He2-157}  
}
\startdata

He$^+$/H$^+$&  0.12    &  0.15    &    \nodata    &  7.99E-02&  4.37E-02&  0.12  
   \\
He$^{+2}$/H$^+$&  9.47E-04\tablenotemark{c}&  3.12E-04\tablenotemark{b}&    
\nodata    &  5.92E-04\tablenotemark{b}&  7.57E-02&  3.99E-04 \\
ICF(He)&      1.00&      1.00&      1.00&      1.00&      1.00&      1.00 \\
O$^o$/H$^+$&  4.96E-06&  1.35E-05&  2.71E-05&  1.64E-05&  7.90E-06&  4.16E-06 \\
O$^+$/H$^+$&  6.48E-05&  2.26E-04&  4.59E-04&  3.19E-04&  4.38E-05&  1.13E-04 \\
O$^{+2}$/H$^+$&  2.82E-04&  4.25E-04&  1.12E-06\tablenotemark{b}&  6.01E-05&  
2.36E-04&  5.59E-05 \\
ICF(O)&      1.01&      1.00&      1.00&      1.01&      2.73&      1.00 \\
N$^{+}$/H$^+$&  1.44E-05&  1.84E-04&  1.98E-04&  1.44E-04&  1.48E-05&  4.84E-05 
\\
ICF(N)&      5.40&      2.89&      1.00&      1.20&     17.46&      1.50 \\
Ne$^{+2}$/H$^+$&  3.55E-05&  1.14E-04&    \nodata    &  
5.11E-06\tablenotemark{c}&  4.43E-05&    \nodata     \\
ICF(Ne)&      1.24&      1.54&   412.40&     6.34&      3.24&      3.02 \\
S$^+$/H$^+$&  1.79E-07&  1.72E-06&  6.47E-06&  1.53E-06&  2.72E-07&  5.09E-07 \\
S$^{+2}$/H$^+_{NIR}$&  2.03E-06&  6.95E-06&  1.77E-06&  6.35E-06&  1.61E-06&  
3.81E-06 \\
S$^{+2}$/H$^+_{6312}$&  1.83E-06&  1.43E-05&    \nodata    &  8.46E-06&  
7.94E-06&  3.05E-06 \\
ICF(S)&      1.21&      1.12&      0.96&      0.99&      1.47&      1.03 \\
Cl$^{+2}$/H$^+$&  9.05E-08\tablenotemark{b}&  3.28E-07\tablenotemark{b}&    
\nodata    &  1.75E-07\tablenotemark{c}&  6.10E-08&  6.50E-08 \\
Cl$^{+3}$/H$^+$&    \nodata    &    \nodata    &    \nodata    &    \nodata    & 
 3.41E-08&    \nodata     \\
ICF(Cl)&      1.01&      1.00&      1.00&      1.01&      2.73&      1.00 \\
Ar$^{+2}$/H$^+$&  1.79E-06&  3.98E-06&    \nodata    &  2.22E-06&  1.17E-07&  
1.15E-06 \\
Ar$^{+3}$/H$^+$&    \nodata    &    \nodata    &    \nodata    &    \nodata    & 
 7.41E-07&    \nodata     \\
ICF(Ar)&      1.24&      \nodata&      \nodata&      \nodata&      2.90&      
\nodata \\
T$_{O3}$(K)&     9100&    6200\tablenotemark{d}&    5800\tablenotemark{d}&    
7000\tablenotemark{d}&    12400&    10500\tablenotemark{d} \\
T$_{N2}$(K)&    11700&     6600&     6000&     7600&     8800&    10300 \\
T$_{O2}$(K)&    12400&     5800&     5900&     7100&     7300&     7400 \\
T$_{S2}$(K)&    16300&     6400&     6300&     6100&    (41800)&        \nodata 
\\
T$_{S3}$(K)&    11500&     7700&        \nodata&     8200&    15500&     9800 \\
N$_{e,S2}$(cm$^{-3}$) &    11700&     2000&     5300&     5300&     1400&     
5200 \\

\enddata

\tablenotetext{a}{Unless otherwise noted, uncertainties in ionic abundances, 
electron temperatures, and electron densities are $\pm$30\%, $\pm$10\%, and 
$\pm$10\%, respectively.}

\tablenotetext{b}{Uncertainty is estimated to be $\pm$50\%}

\tablenotetext{c}{Uncertainty is estimated to be $\pm$75\%}
\tablenotetext{d}{T$_{O3}$ estimated from T$_{N2}$; see eq.~2 and discussion}

\end{deluxetable}

\begin{deluxetable}{lrrrrrr}
\tabletypesize{\scriptsize}
\setlength{\topmargin}{-1in}
\setlength{\tabcolsep}{0.03in}
\tablecolumns{7}
\tablewidth{0in}
\tablenum{2C}
\tablecaption{Ionic Abundances, Temperatures, \& Densities\tablenotemark{a}}
\tablehead{
\colhead{Parameter} & 
\colhead{He2-158} &
\colhead{IC 1297} &
\colhead{IC 2448} &
\colhead{IC 2501} &
\colhead{IC 2621} &
\colhead{IC 4776}  
}
\startdata

He$^+$/H$^+$&  0.12    &  0.10    &  7.78E-02&  3.71E-03&  8.24E-02&  0.11     
\\
He$^{+2}$/H$^+$&   \nodata    &  3.14E-02&  3.75E-02&    \nodata    &  3.73E-02& 
   \nodata     \\
ICF(He)&      1.00&      1.00&      1.00&      1.00&      1.00&      1.00 \\
O$^o$/H$^+$&  1.55E-05&  6.16E-06&    \nodata    &  7.88E-06&  9.46E-06&  
4.31E-06 \\
O$^+$/H$^+$&  8.04E-05&  4.34E-05&  2.54E-07\tablenotemark{c}&  3.41E-05&  
1.69E-05&  1.48E-05 \\
O$^{+2}$/H$^+$&  2.43E-04&  5.04E-04&  2.19E-04&  3.91E-04&  2.96E-04&  4.43E-04 
\\
ICF(O)&      1.00&      1.31&      1.48&      1.00&      1.45&      1.00 \\
N$^{+}$/H$^+$&  2.50E-05&  1.36E-05&  7.25E-08&  1.14E-05&  1.84E-05&  4.84E-06 
\\
ICF(N)&      4.02&     16.60&   1279.69&     12.47&     26.92&     30.97 \\
Ne$^{+2}$/H$^+$&  5.39E-05&  1.23E-04&  4.61E-05&  1.03E-04&  6.73E-05&  
9.84E-05 \\
ICF(Ne)&      1.33&      1.43&      1.48&      1.09&      1.54&      1.03 \\
S$^+$/H$^+$&  5.47E-07&  5.57E-07&  1.82E-09&  2.13E-07&  2.65E-07&  1.62E-07 \\
S$^{+2}$/H$^+_{NIR}$&  3.00E-06&  4.16E-06&  2.32E-07\tablenotemark{d}&  
1.60E-06&  3.53E-06&  3.57E-06 \\
S$^{+2}$/H$^+_{6312}$&  2.66E-06&  8.48E-06\tablenotemark{b}&  1.96E-07&  
2.19E-06&  3.02E-06&  2.30E-06 \\
ICF(S)&      1.16&      1.45&     39.18&      1.37&      1.64&      1.72 \\
Cl$^{+2}$/H$^+$&  7.41E-08&  1.02E-07&  1.51E-08&  8.65E-08&  6.13E-08&  
9.22E-08 \\
Cl$^{+3}$/H$^+$&  8.33E-09\tablenotemark{c}&  4.77E-08&  3.17E-08&  8.45E-09&  
4.34E-08&  1.46E-08 \\
ICF(Cl)&      1.00&      1.31&      1.48&      1.00&      1.45&      1.00 \\
Ar$^{+2}$/H$^+$&  1.11E-06&  1.59E-06&  2.74E-07&  1.49E-06&  1.72E-06&  
1.51E-06 \\
Ar$^{+3}$/H$^+$&    \nodata    &  3.21E-07&  5.64E-07&  1.24E-07&  9.00E-07&  
8.28E-08\tablenotemark{c} \\
ICF(Ar)&      1.33&      1.40&      1.48&      1.09&      1.51&      1.03 \\
T$_{O3}$(K)&     9200&     9900&    12500&     9500&    12500&     9500 \\
T$_{N2}$(K)&     10000&     8900&    (22400)&    11200&    13400&    13500 \\
T$_{O2}$(K)&     7800&     8100&        \nodata&    10600&    (23500)&    15600 
\\
T$_{S2}$(K)&    15200&    (19300)&        \nodata&    11700&    (25000)&    
15000 \\
T$_{S3}$(K)&     9700&    11000&    (25200)&    12700&    12800&    11500 \\
N$_{e,S2}$(cm$^{-3}$) &     2400&     2400&       10&     4800&     8300&     
4000 \\

\enddata

\tablenotetext{a}{Unless otherwise noted, uncertainties in ionic abundances, 
electron temperatures, and electron densities are $\pm$30\%, $\pm$10\%, and 
$\pm$10\%, respectively.}

\tablenotetext{b}{Uncertainty is estimated to be $\pm$50\%}

\tablenotetext{c}{Uncertainty is estimated to be $\pm$75\%}
\tablenotetext{d}{S$^{+2}$ abundance calculated using T$_{O3}$}

\end{deluxetable}

\begin{deluxetable}{lrrrrrr}
\tabletypesize{\scriptsize}
\setlength{\topmargin}{-1in}
\setlength{\tabcolsep}{0.03in}
\tablecolumns{7}
\tablewidth{0in}
\tablenum{2D}
\tablecaption{Ionic Abundances, Temperatures, \& Densities\tablenotemark{a}}
\tablehead{
\colhead{Parameter} & 
\colhead{J320} &
\colhead{M1-5} &
\colhead{M1-25} &
\colhead{M1-34} &
\colhead{M1-38} &
\colhead{M1-50}  
}
\startdata

He$^+$/H$^+$&  0.11    &  0.11    &  0.14    &  0.14    &    \nodata    &  
9.71E-02 \\
He$^{+2}$/H$^+$&  3.19E-03&  1.91E-04\tablenotemark{c}&  
6.07E-03\tablenotemark{c}&  1.35E-02&    \nodata    &  2.09E-02 \\
ICF(He)&      1.00&      1.00&      1.00&      1.00&      1.00&      1.00 \\
O$^o$/H$^+$&  1.36E-06&  3.06E-06&  9.76E-06&  9.56E-05&  2.38E-05&  7.34E-07 \\
O$^+$/H$^+$&  4.66E-06&  2.55E-05&  1.69E-04&  2.88E-04&  5.88E-04&  7.68E-06 \\
O$^{+2}$/H$^+$&  2.37E-04&  1.04E-04&  3.14E-04&  3.32E-04&  
2.43E-06\tablenotemark{b}&  5.42E-04 \\
ICF(O)&      1.03&      1.00&      1.04&      1.10&      1.00&      1.22 \\
N$^{+}$/H$^+$&  2.45E-06&  1.49E-05&  7.44E-05&  1.91E-04&  1.25E-04&  1.77E-06 
\\
ICF(N)&     53.42&      5.07&      2.98&      2.36&      1.00&     87.03 \\
Ne$^{+2}$/H$^+$&  5.06E-05&  9.47E-06&  2.09E-05&  1.33E-04&  
1.64E-05\tablenotemark{c}&  1.18E-04 \\
ICF(Ne)&      1.05&      1.25&      1.60&      2.05&   243.25&      1.23 \\
S$^+$/H$^+$&  7.65E-08&  1.33E-07&  1.03E-06&  4.68E-06&  4.01E-06&  8.70E-08 \\
S$^{+2}$/H$^+_{NIR}$&  5.67E-07&  2.04E-06&  7.20E-06&  6.94E-06&  1.70E-06&  
1.55E-06 \\
S$^{+2}$/H$^+_{6312}$&  2.31E-06&  1.21E-06&  1.12E-05&  1.02E-05&  
4.71E-06\tablenotemark{c}&  3.48E-06 \\
ICF(S)&      2.10&      1.20&      1.12&      1.09&      0.96&      2.66 \\
Cl$^{+2}$/H$^+$&  2.44E-08&  4.93E-08\tablenotemark{b}&  1.92E-07&  2.05E-07&    
\nodata    &  7.27E-08\tablenotemark{b} \\
Cl$^{+3}$/H$^+$&  4.47E-08&    \nodata    &    \nodata    &    \nodata    &    
\nodata    &  8.40E-08 \\
ICF(Cl)&      1.03&      1.00&      1.04&      1.10&      1.00&      1.22 \\
Ar$^{+2}$/H$^+$&  4.90E-07&  7.86E-07&  3.34E-06&  2.93E-06&  
1.90E-07\tablenotemark{c}&  1.00E-06 \\
Ar$^{+3}$/H$^+$&  4.04E-07&  1.86E-08\tablenotemark{c}&    \nodata    &  
1.36E-07\tablenotemark{b}&    \nodata    &  8.86E-07 \\
ICF(Ar)&      1.05&      1.25&     \nodata&     \nodata&     \nodata&      1.23 
\\
T$_{O3}$(K)&    12100&    10000&     8400&     9200&    6300\tablenotemark{d}&   
 10300 \\
T$_{N2}$(K)&     9900.&    11800&     8200&     8200&     6700&    10200 \\
T$_{O2}$(K)&    16100&    (29500)&     7900.&     5400&     5900&    11700 \\
T$_{S2}$(K)&     6100&    11700&     9100&     9400&     6600&    (25300) \\
T$_{S3}$(K)&    17600&    10100&     9300&     9100&     8500&    13900 \\
N$_{e,S2}$(cm$^{-3}$) &     4600&     6400&     4300&      700&     5100&     
4500 \\

\enddata

\tablenotetext{a}{Unless otherwise noted, uncertainties in ionic abundances, 
electron temperatures, and electron densities are $\pm$30\%, $\pm$10\%, and 
$\pm$10\%, respectively.}

\tablenotetext{b}{Uncertainty is estimated to be $\pm$50\%}

\tablenotetext{c}{Uncertainty is estimated to be $\pm$75\%}
\tablenotetext{d}{T$_{O3}$ estimated from T$_{N2}$; see eq.~2 and discussion}

\end{deluxetable}
\begin{deluxetable}{lrrrrrr}
\tabletypesize{\scriptsize}
\setlength{\topmargin}{-1in}
\setlength{\tabcolsep}{0.03in}
\tablecolumns{7}
\tablewidth{0in}
\tablenum{2E}
\tablecaption{Ionic Abundances, Temperatures, \& Densities\tablenotemark{a}}
\tablehead{
\colhead{Parameter} & 
\colhead{M1-54} &
\colhead{M1-57} &
\colhead{M2-10} &
\colhead{M3-4} &
\colhead{M3-6} &
\colhead{M3-15\tablenotemark{f}}  
}
\startdata

He$^+$/H$^+$&  0.14    &  7.91E-02    &  0.13    &  0.12    &  0.12    &  0.12   
  \\
He$^{+2}$/H$^+$&  1.63E-02&  4.69E-02&    \nodata    &  3.01E-02&  9.85E-04&  
3.51E-03\tablenotemark{c} \\
ICF(He)&      1.00&      1.00&      1.00&      1.00&      1.00&      1.00 \\
O$^o$/H$^+$&  6.29E-05&  3.48E-05&  1.94E-05&  4.12E-05&    \nodata    &  
3.20E-06 \\
O$^+$/H$^+$&  1.98E-04&  8.30E-05&  3.04E-04&  1.54E-04&  1.45E-05&  1.97E-05 \\
O$^{+2}$/H$^+$&  3.30E-04&  3.19E-04&  3.01E-04&  2.57E-04&  5.39E-04&  6.43E-04 
\\
ICF(O)&      1.12&      1.59&      1.00&      1.26&      1.01&      1.03 \\
N$^{+}$/H$^+$&  1.91E-04&  5.69E-05    &  1.51E-04&  5.75E-05&  1.52E-06&  
9.77E-06 \\
ICF(N)&      2.99&     7.71&      1.99&      3.35&     38.41&     34.60 \\
Ne$^{+2}$/H$^+$&  1.35E-04&  7.31E-05&  7.56E-05&  8.75E-05&  1.31E-04&  
1.51E-04 \\
ICF(Ne)&      1.79&      2.01&      2.01&      2.01&      1.04&      1.06 \\
S$^+$/H$^+$&  3.61E-06&  1.76E-06&  1.93E-06&  4.41E-07&  3.77E-08&  3.21E-07 \\
S$^{+2}$/H$^+_{NIR}$&  6.29E-06&  4.08E-06\tablenotemark{d}&  7.12E-06&  
9.87E-07&  3.81E-06&  4.78E-06 \\
S$^{+2}$/H$^+_{6312}$&  7.77E-06&  1.08E-05&  9.81E-06&  
1.73E-06\tablenotemark{c}&  3.20E-06&  3.76E-06 \\
ICF(S)&      1.12&      1.27&      1.07&      1.14&      1.84&      1.78 \\
Cl$^{+2}$/H$^+$&  1.21E-07&  1.01E-07&  2.17E-07\tablenotemark{b}&  
6.66E-08\tablenotemark{b}&  9.43E-08&  2.18E-07 \\
Cl$^{+3}$/H$^+$&  4.15E-08\tablenotemark{b}&  5.42E-08&  
5.74E-08\tablenotemark{c}&    \nodata    &  2.25E-08\tablenotemark{c}&  1.86E-08 
\\
ICF(Cl)&      1.12&      1.59&      1.00&      1.26&      1.01&      1.03 \\
Ar$^{+2}$/H$^+$&  2.51E-06&  1.76E-06&  3.45E-06&  1.43E-06\tablenotemark{b}&  
2.81E-06&  2.68E-06 \\
Ar$^{+3}$/H$^+$&  4.92E-07&  7.99E-07&    \nodata    &  2.71E-07&  1.44E-07&  
3.43E-07\tablenotemark{c} \\
ICF(Ar)&      \nodata&1.83       &      \nodata&      \nodata&      1.04&      
1.06 \\
T$_{O3}$(K)&     9800&    12700&    6300\tablenotemark{e}&    11500&     8000&   
  8200 \\
T$_{N2}$(K)&     8700&    11300&     6700&     9300&     8700&    (16400) \\
T$_{O2}$(K)&     5900&     9500&     5100&    14000&     9500&    10800 \\
T$_{S2}$(K)&    11200&     7600&    11400&     8500&        \nodata&     9000 \\
T$_{S3}$(K)&     9200&    (20500)&     7200&    11000&     8400&    10100 \\
N$_{e,S2}$(cm$^{-3}$) &     1500&     3600&     1300&      200&     1700&     
3700 \\

\enddata

\tablenotetext{a}{Unless otherwise noted, uncertainties in ionic abundances, 
electron temperatures, and electron densities are $\pm$30\%, $\pm$10\%, and 
$\pm$10\%, respectively.}

\tablenotetext{b}{Uncertainty is estimated to be $\pm$50\%}

\tablenotetext{c}{Uncertainty is estimated to be $\pm$75\%}
\tablenotetext{d}{S$^{+2}$ abundance calculated using T$_{O3}$}
\tablenotetext{e}{T$_{O3}$ estimated from T$_{N2}$; see eq.~2 and discussion}
\tablenotetext{f}{Abundances of low ionization species were calculated using 
T$_{O2}$ rather than T$_{N2}$, due to the
unusually high value of the latter}

\end{deluxetable}
\begin{deluxetable}{lrrrrrr}
\tabletypesize{\scriptsize}
\setlength{\topmargin}{-1in}
\setlength{\tabcolsep}{0.03in}
\tablecolumns{7}
\tablewidth{0in}
\tablenum{2F}
\tablecaption{Ionic Abundances, Temperatures, \& Densities\tablenotemark{a}}
\tablehead{
\colhead{Parameter} & 
\colhead{NGC 2792} &
\colhead{NGC 2867} &
\colhead{NGC 3195} &
\colhead{NGC 3211} &
\colhead{NGC 3242A} &
\colhead{NGC 3242B}  
}
\startdata

He$^+$/H$^+$&  1.89E-02&  8.89E-02&  0.12    &  3.07E-02&  8.16E-02&  9.24E-02 
\\
He$^{+2}$/H$^+$&  8.80E-02&  3.04E-02&  1.14E-02&  8.09E-02&  2.87E-02&  
1.56E-02 \\
ICF(He)&      1.00&      1.00&      1.00&      1.00&      1.00&      1.00 \\
O$^o$/H$^+$&    \nodata    &  1.29E-05&  5.49E-05&    \nodata    &    \nodata    
&    \nodata     \\
O$^+$/H$^+$&  2.81E-06\tablenotemark{b}&  4.57E-05&  2.86E-04&  8.83E-06&  
2.01E-06&  6.46E-06\tablenotemark{b} \\
O$^{+2}$/H$^+$&  1.33E-04&  3.44E-04&  3.43E-04&  2.22E-04&  3.21E-04&  3.22E-04 
\\
ICF(O)&      5.65&      1.34&      1.09&      3.64&      1.35&      1.17 \\
N$^{+}$/H$^+$&  5.12E-07\tablenotemark{c}&  1.21E-05&  1.28E-04&  1.72E-06&  
2.33E-07\tablenotemark{c}&  7.94E-07\tablenotemark{b} \\
ICF(N)&    272.18&     11.43&      2.41&     94.82&    217.77&     59.35 \\
Ne$^{+2}$/H$^+$&  2.18E-05&  7.19E-05&  1.38E-04&  3.47E-05&  6.60E-05&  
7.04E-05 \\
ICF(Ne)&      5.77&      1.52&      2.01&      3.78&      1.36&      1.19 \\
S$^+$/H$^+$&  2.71E-08&  3.74E-07&  3.53E-06&  1.13E-07&  1.39E-08&  2.56E-08 \\
S$^{+2}$/H$^+_{NIR}$&  6.18E-07\tablenotemark{d}&  2.11E-06&  5.74E-06&  
1.47E-06&  5.52E-07&  7.74E-07 \\
S$^{+2}$/H$^+_{6312}$&  3.06E-06&  4.13E-06&  9.16E-06&  5.41E-06&  1.02E-06&  
2.12E-06 \\
ICF(S)&      6.02&      1.35&      1.09&      2.79&      4.97&      2.20 \\
Cl$^{+2}$/H$^+$&  2.57E-08&  7.31E-08&  1.34E-07&  5.56E-08&  2.71E-08&  
3.10E-08 \\
Cl$^{+3}$/H$^+$&  6.58E-08&  3.68E-08&  1.15E-08&  6.87E-08&  4.87E-08&  
6.35E-08 \\
ICF(Cl)&      5.65&      1.34&      1.09&      3.64&      1.35&      1.17 \\
Ar$^{+2}$/H$^+$&  4.25E-07&  1.04E-06&  2.74E-06&  8.63E-07&  5.16E-07&  
7.28E-07 \\
Ar$^{+3}$/H$^+$&  7.08E-07&  2.65E-07&  1.45E-07&  8.61E-07&  5.90E-07&  
5.37E-07 \\
ICF(Ar)&      5.67&      1.47&      \nodata&      3.67&      1.36&      1.19 \\
T$_{O3}$(K)&    13700&    11200&     8900&    13300&    11200&    11200 \\
T$_{N2}$(K)&    10200&    10100&     8000&    10300&    12100&     9200 \\
T$_{O2}$(K)&     7900&     8200&    11300&     6700&     9600&     9300 \\
T$_{S2}$(K)&        \nodata&    (27700)&    12500&    (67900)&        \nodata&   
     \nodata \\
T$_{S3}$(K)&    (26200)&    12800&     9000&    (20100)&    15800&    12800 \\
N$_{e,S2}$(cm$^{-3}$) &     2800&     2100&      200&     1200&     2100&     
1200 \\

\enddata

\tablenotetext{a}{Unless otherwise noted, uncertainties in ionic abundances, 
electron temperatures, and electron densities are $\pm$30\%, $\pm$10\%, and 
$\pm$10\%, respectively.}

\tablenotetext{b}{Uncertainty is estimated to be $\pm$50\%}

\tablenotetext{c}{Uncertainty is estimated to be $\pm$75\%}
\tablenotetext{d}{S$^{+2}$ abundance calculated using T$_{O3}$}

\end{deluxetable}
\begin{deluxetable}{lrrrrrr}
\tabletypesize{\scriptsize}
\setlength{\topmargin}{-1in}
\setlength{\tabcolsep}{0.03in}
\tablecolumns{7}
\tablewidth{0in}
\tablenum{2G}
\tablecaption{Ionic Abundances, Temperatures, \& Densities\tablenotemark{a}}
\tablehead{
\colhead{Parameter} & 
\colhead{NGC 5307} &
\colhead{NGC 6309} &
\colhead{NGC 6439} &
\colhead{NGC 6563} &
\colhead{NGC 6565} &
\colhead{NGC 6629}  
}
\startdata

He$^+$/H$^+$&  6.34E-02&  4.53E-02&  0.12    &  0.11    &  0.11    &  0.11     
\\
He$^{+2}$/H$^+$&  3.83E-02&  7.50E-02&  2.02E-02&  1.48E-02&  1.43E-02&  
1.07E-03\tablenotemark{b} \\
ICF(He)&      1.00&      1.00&      1.00&      1.00&      1.00&      1.00 \\
O$^o$/H$^+$&  2.42E-07\tablenotemark{c}&  5.12E-06&  8.96E-06&  5.57E-05&  
5.40E-05&    \nodata     \\
O$^+$/H$^+$&  3.31E-06\tablenotemark{b}&  2.47E-05&  4.03E-05&  2.33E-04&  
2.31E-04&  1.70E-05 \\
O$^{+2}$/H$^+$&  2.37E-04&  2.24E-04&  5.11E-04&  3.28E-04&  4.09E-04&  4.29E-04 
\\
ICF(O)&      1.60&      2.66&      1.17&      1.13&      1.13&      1.01 \\
N$^{+}$/H$^+$&  4.41E-07&  6.22E-06&  2.87E-05&  7.34E-05&  9.41E-05&  2.35E-06 
\\
ICF(N)&    116.24&     26.76&     16.01&      2.73&      3.14&     26.55 \\
Ne$^{+2}$/H$^+$&  5.52E-05&  5.38E-05&  1.49E-04&  1.10E-04&  1.36E-04&  
8.74E-05 \\
ICF(Ne)&      1.63&      2.95&      1.26&      1.94&      1.77&      1.05 \\
S$^+$/H$^+$&  2.89E-08\tablenotemark{c}&  3.11E-07&  7.16E-07&  9.80E-07&  
2.91E-06&  3.82E-08 \\
S$^{+2}$/H$^+_{NIR}$&  5.29E-07\tablenotemark{d}&  2.77E-06&  5.24E-06&  
1.99E-06&  5.66E-06&  2.01E-06 \\
S$^{+2}$/H$^+_{6312}$&  1.48E-06\tablenotemark{c}&  6.71E-06&  8.11E-06&    
\nodata    &  7.77E-06&  1.35E-06 \\
ICF(S)&      3.16&      1.64&      1.44&      1.11&      1.13&      1.64 \\
Cl$^{+2}$/H$^+$&  1.27E-08\tablenotemark{b}&  4.32E-08&  1.48E-07&  
2.62E-08\tablenotemark{c}&  1.52E-07&  9.42E-08 \\
Cl$^{+3}$/H$^+$&  2.98E-08&  8.54E-08&  7.66E-08\tablenotemark{b}&    \nodata    
&  1.33E-08\tablenotemark{c}&    \nodata     \\
ICF(Cl)&      1.60&      2.66&      1.17&      1.13&      1.13&      1.01 \\
Ar$^{+2}$/H$^+$&  2.92E-07&  9.82E-07&  2.79E-06&  1.89E-06&  2.37E-06&  
1.79E-06 \\
Ar$^{+3}$/H$^+$&  5.46E-07&  1.08E-06&  1.02E-06&    \nodata    &  2.91E-07&    
\nodata     \\
ICF(Ar)&      1.62&      2.76&      1.25&      \nodata&      \nodata&      1.05 
\\
T$_{O3}$(K)&    12300&    11700&     9700&    10300&    10100&     8500 \\
T$_{N2}$(K)&    11000&     8200&     9600&     9000&     9400&    10300 \\
T$_{O2}$(K)&     9900&     9500&     8200&    13100&     6100&     6500 \\
T$_{S2}$(K)&        \nodata&     9800&    (33600)&        \nodata&    11800&     
   \nodata \\
T$_{S3}$(K)&    (20300)&    10500&    11100&        \nodata&    10300&     9200 
\\
N$_{e,S2}$(cm$^{-3}$)&     2600&     3200&     3000&      100&     1300&     
1100 \\ 

\enddata

\tablenotetext{a}{Unless otherwise noted, uncertainties in ionic abundances, 
electron temperatures, and electron densities are $\pm$30\%, $\pm$10\%, and 
$\pm$10\%, respectively.}

\tablenotetext{b}{Uncertainty is estimated to be $\pm$50\%}

\tablenotetext{c}{Uncertainty is estimated to be $\pm$75\%}
\tablenotetext{d}{S$^{+2}$ abundance calculated using T$_{O3}$}

\end{deluxetable}

\begin{deluxetable}{lrrrr}
\tabletypesize{\scriptsize}
\setlength{\topmargin}{-1in}
\setlength{\tabcolsep}{0.03in}
\tablecolumns{5}
\tablewidth{0in}
\tablenum{2H}
\tablecaption{Ionic Abundances, Temperatures, \& Densities\tablenotemark{a}}
\tablehead{
\colhead{Parameter} & 
\colhead{PB6\tablenotemark{d}} &
\colhead{PC14} &
\colhead{Pe1-18} &
\colhead{Th2-A}  
}
\startdata

He$^+$/H$^+$&    3.66E-02    &  0.12    &  0.15    &  8.13E-02 \\
He$^{+2}$/H$^+$&    0.14    &  3.09E-03\tablenotemark{b}&  
4.45E-04\tablenotemark{c}&  4.67E-02 \\
ICF(He)&      1.00&      1.00&      1.00&      1.00 \\
O$^o$/H$^+$&  5.69E-06    &  9.97E-06&  4.10E-06&  4.42E-06\tablenotemark{b} \\
O$^+$/H$^+$&  2.94E-05&  7.73E-05&  7.01E-06\tablenotemark{c}&  3.41E-05 \\
O$^{+2}$/H$^+$&  1.13E-04    &  6.90E-04&  4.36E-04&  3.62E-04 \\
ICF(O)&      4.74&      1.03&      1.00&      1.57 \\
N$^{+}$/H$^+$&  3.32E-05&  1.37E-05&  1.45E-05&  1.66E-05 \\
ICF(N)&      23.01&     10.19&     63.34&     18.26 \\
Ne$^{+2}$/H$^+$&  2.68E-05    &  1.89E-04&  1.18E-04&  1.13E-04 \\
ICF(Ne)&      5.98&      1.14&      1.02&      1.72 \\
S$^+$/H$^+$&  3.11E-07&  7.16E-07&  2.95E-07&  3.10E-07 \\
S$^{+2}$/H$^+_{NIR}$&  1.10E-06&  4.88E-06&  4.31E-06&  
2.44E-06\tablenotemark{b,e} \\
S$^{+2}$/H$^+_{6312}$&  4.16E-06    &  1.12E-05&  2.59E-06&  
5.79E-06\tablenotemark{b} \\
ICF(S)&      1.57&      1.32&      2.27&      1.48 \\
Cl$^{+2}$/H$^+$&    \nodata    &  1.59E-07\tablenotemark{c}&  
9.72E-08\tablenotemark{c}&    \nodata     \\
Cl$^{+3}$/H$^+$&  3.82E-08&  4.34E-08&  3.02E-08&  4.92E-08\tablenotemark{c} \\
ICF(Cl)&      4.74&      1.03&      1.00&      1.57 \\
Ar$^{+2}$/H$^+$&  5.97E-07&  2.23E-06&  3.15E-06&  1.63E-06 \\
Ar$^{+3}$/H$^+$&  5.84E-07    &  4.10E-07&  3.54E-07\tablenotemark{b}&  
3.22E-07\tablenotemark{c} \\
ICF(Ar)&      4.96&      1.14&      1.02&      1.67 \\
T$_{O3}$(K)&    14600&     8800&     9800&    11600 \\
T$_{N2}$(K)&     9800&     8200&    15300&    11700 \\
T$_{O2}$(K)&     7600&     7400&    (34900)&     5900 \\
T$_{S2}$(K)&        \nodata&     6900&    (33400)&        \nodata \\
T$_{S3}$(K)&    16300&    10200&    12500&    (21300) \\
N$_{e,S2}$(cm$^{-3}$) &     2200&     2200&    12100&     1200 \\

\enddata

\tablenotetext{a}{Unless otherwise noted, uncertainties in ionic abundances, 
electron temperatures, and electron densities are $\pm$30\%, $\pm$10\%, and 
$\pm$10\%, respectively.}

\tablenotetext{b}{Uncertainty is estimated to be $\pm$50\%}

\tablenotetext{c}{Uncertainty is estimated to be $\pm$75\%}
\tablenotetext{d}{Our observations of PB6 cover only 5800-9600~{\AA};
blue line intensities were taken from Kaler (1991)and were merged with
ours to calculate abundances.} 
\tablenotetext{e}{S$^{+2}$ abundance
calculated using T$_{O3}$}

\end{deluxetable}

\begin{deluxetable}{lcccccccc}
\tabletypesize{\scriptsize}
\setlength{\tabcolsep}{0.03in}
\tablecolumns{9}
\tablewidth{0in}
\tablenum{3A}
\tablecaption{Elemental Abundances\tablenotemark{a}}
\tablehead{
\colhead{Element} & 
\colhead{Cn2-1} &
\colhead{Fg 1} &
\colhead{He2-21} &
\colhead{He2-37} &
\colhead{He2-48} &
\colhead{He2-55}   &
\colhead{Sun\tablenotemark{b}} &
\colhead{Orion\tablenotemark{c}}
}
\startdata

He/H&      0.13&      0.13&      0.12&      0.12&      0.11&      0.13&      
0.10&      0.10 \\
O/H ($\times 10^4$)&      7.14&      3.73&      2.99&     11.00&      4.23&      
6.87&      7.41&      5.25 \\
N/H ($\times 10^4$)&      2.78&      1.60&      0.44&      3.52&      1.32&      
1.40&      0.93&      0.60 \\
Ne/H ($\times 10^4$)&      1.65&      1.10&      0.49&      2.48&      1.14&     
 1.78&      1.20&      0.78 \\
S/H ($\times 10^5$)&      0.62&      0.46&      0.09&      0.45&      0.18&      
0.35&      2.14&      1.48 \\
Cl/H ($\times 10^7$)&      1.78&      1.24&      0.83&      2.98&      0.82&     
 3.62&      3.16&      2.14 \\
Ar/H ($\times 10^6$)&      2.72&      2.13&      1.09&      4.75&      
1.07\tablenotemark{d}&      5.04&      3.31&      3.09 \\
N/O&      0.39&      0.43&      0.15&      0.32&      0.31&      0.20&      
0.13&      0.11 \\ 
Ne/O&      0.23&      0.29&      0.16&      0.23&      0.27&      0.26&      
0.16&      0.15 \\
S/O ($\times 10^1$)&      0.09&      0.12&      0.03&      0.04&      0.04&      
0.05&      0.29&      0.28 \\
Cl/O ($\times 10^3$) &      0.25&      0.33&      0.28&      0.27&      0.19&    
  0.53&      0.43&      0.41 \\
Ar/O ($\times 10^2$)&      0.38&      0.57&      0.36&      0.43&      
0.25\tablenotemark{d}&      0.73&      0.45&      0.59 \\  

\enddata

\tablenotetext{a} {Uncertainties in elemental abundances are generally as 
follows. S/O: $\pm$30\%; Cl/O: $\pm$50\%; Ar/O: $\pm$75\%.}
\tablenotetext{b}{Grevesse et al. (1996)}
\tablenotetext{c}{Esteban et al. (1998), Table 19, gas + dust}
\tablenotetext{d}{Ar=Ar$^{+2}$}

\end{deluxetable}

\begin{deluxetable}{lcccccccc}
\tabletypesize{\scriptsize}
\setlength{\tabcolsep}{0.03in}
\tablecolumns{9}
\tablewidth{0in}
\tablenum{3B}
\tablecaption{Elemental Abundances\tablenotemark{a}}
\tablehead{
\colhead{Element} & 
\colhead{He2-115} &
\colhead{He2-123} &
\colhead{He2-138} &
\colhead{He2-140} &
\colhead{He2-141} &
\colhead{He2-157}   &
\colhead{Sun\tablenotemark{b}} &
\colhead{Orion\tablenotemark{c}}
}
\startdata

He/H&      0.12&      0.15&      0.00&      0.08&      0.12&      0.12&      
0.10&      0.10 \\
O/H ($\times 10^4$)&      3.50&      6.52&      4.60&      3.82&      7.64&      
1.69&      7.41&      5.25 \\
N/H ($\times 10^4$)&      0.78&      5.31&      1.98&      1.72&      2.58&      
0.73&      0.93&      0.60 \\
Ne/H ($\times 10^4$)&      0.44&      1.75&      \nodata&      0.32&      1.44&  
    \nodata&      1.20&      0.78 \\
S/H ($\times 10^5$)&      0.27&      0.97&      0.79&      0.78&      0.28&      
0.44&      2.14&      1.48 \\
Cl/H ($\times 10^7$)&      0.91&      3.29&      \nodata&      1.76&      2.60&  
    0.65&      3.16&      2.14 \\
Ar/H ($\times 10^6$)&      2.22&      3.98\tablenotemark{d}&      \nodata&      
2.22\tablenotemark{d}&      2.49&      1.15\tablenotemark{d}&      3.31&      
3.09 \\
N/O &      0.22&      0.81&      0.43&      0.45&      0.34&      0.43&      
0.13&      0.11 \\
Ne/O&      0.13&      0.27&      \nodata&      0.09&      0.19&      \nodata&    
  0.16&      0.15 \\
S/O ($\times 10^1$)&      0.08&      0.15&      0.17&      0.20&      0.04&      
0.26&      0.29&      0.28 \\
Cl/O ($\times 10^3$)&      0.26&      0.50&      \nodata&      0.46&      0.34&  
    0.39&      0.43&      0.41 \\
Ar/O ($\times 10^2$) &      0.63&      0.61\tablenotemark{d}&      \nodata&      
0.58\tablenotemark{d}&      0.33&      0.68\tablenotemark{d}&      0.45&      
0.59 \\ 

\enddata

\tablenotetext{a} {Uncertainties in elemental abundances are generally as 
follows. S/O: $\pm$30\%; Cl/O: $\pm$50\%; Ar/O: $\pm$75\%.}
\tablenotetext{b}{Grevesse et al. (1996)}
\tablenotetext{c}{Esteban et al. (1998), Table 19, gas + dust}
\tablenotetext{d}{Ar=Ar$^{+2}$}

\end{deluxetable}

\begin{deluxetable}{lcccccccc}
\tabletypesize{\scriptsize}
\setlength{\tabcolsep}{0.03in}
\tablecolumns{9}
\tablewidth{0in}
\tablenum{3C}
\tablecaption{Elemental Abundances\tablenotemark{a}}
\tablehead{
\colhead{Element} & 
\colhead{He2-158} &
\colhead{IC 1297} &
\colhead{IC 2448} &
\colhead{IC 2501} &
\colhead{IC 2621} &
\colhead{IC 4776}   &
\colhead{Sun\tablenotemark{b}} &
\colhead{Orion\tablenotemark{c}}
}
\startdata

He/H&      0.12&      0.13&      0.12&      \nodata&      0.12&      0.11&      
0.10&      0.10 \\
O/H ($\times 10^4$)&      3.24&      7.20&      3.25&      4.25&      4.55&      
4.58&      7.41&      5.25 \\
N/H ($\times 10^4$)&      1.01&      2.26&      0.93&      1.43&      4.96&      
1.50&      0.93&      0.60 \\
Ne/H ($\times 10^4$)&      0.72&      1.76&      0.68&      1.12&      1.03&     
 1.02&      1.20&      0.78 \\
S/H ($\times 10^5$)&      0.41&      0.68&      0.92&      0.25&      0.62&      
0.64&      2.14&      1.48 \\
Cl/H ($\times 10^7$)&      0.82&      1.97&      0.69&      0.95&      1.52&     
 1.07&      3.16&      2.14 \\
Ar/H ($\times 10^6$)&      1.48&      2.68&      1.24&      1.75&      3.96&     
 1.65&      3.31&      3.09 \\
N/O &      0.31&      0.31&      0.29&      0.34&      1.09&      0.33&      
0.13&      0.11 \\
Ne/O&      0.22&      0.24&      0.21&      0.26&      0.23&      0.22&      
0.16&      0.15 \\
S/O ($\times 10^1$)&      0.13&      0.10&      0.28&      0.06&      0.14&      
0.14&      0.29&      0.28 \\
Cl/O ($\times 10^3$)&      0.25&      0.27&      0.21&      0.22&      0.33&     
 0.23&      0.43&      0.41 \\
Ar/O ($\times 10^2$)&      0.46&      0.37&      0.38&      0.41&      0.87&     
 0.36&      0.45&      0.59 \\

\enddata

\tablenotetext{a} {Uncertainties in elemental abundances are generally as 
follows. S/O: $\pm$30\%; Cl/O: $\pm$50\%; Ar/O: $\pm$75\%.}
\tablenotetext{b}{Grevesse et al. (1996)}
\tablenotetext{c}{Esteban et al. (1998), Table 19, gas + dust}

\end{deluxetable}
\begin{deluxetable}{lcccccccc}
\tabletypesize{\scriptsize}
\setlength{\tabcolsep}{0.03in}
\tablecolumns{9}
\tablewidth{0in}
\tablenum{3D}
\tablecaption{Elemental Abundances\tablenotemark{a}}
\tablehead{
\colhead{Element} & 
\colhead{J320} &
\colhead{M1-5} &
\colhead{M1-25} &
\colhead{M1-34} &
\colhead{M1-38} &
\colhead{M1-50}   &
\colhead{Sun\tablenotemark{b}} &
\colhead{Orion\tablenotemark{c}}
}
\startdata

He/H&      0.11&      0.11&      0.15&      0.15&      \nodata&      0.12&      
0.10&      0.10 \\
O/H ($\times 10^4$)&      2.49&      1.29&      5.03&      6.79&      5.91&      
6.68&      7.41&      5.25 \\
N/H ($\times 10^4$)&      1.31&      0.76&      2.22&      4.51&      1.25&      
1.54&      0.93&      0.60 \\
Ne/H ($\times 10^4$)&      0.53&      0.12&      0.33&      2.72&     39.92&     
 1.45&      1.20&      0.78 \\
S/H ($\times 10^5$)&      0.14&      0.26&      0.92&      1.27&      0.55&      
0.44&      2.14&      1.48 \\
Cl/H ($\times 10^7$)&      0.71&      0.49&      2.00&      2.24&      \nodata&  
    1.90&      3.16&      2.14 \\
Ar/H ($\times 10^6$)&      0.94&      1.00&      3.34\tablenotemark{d}&      
2.93\tablenotemark{d}&      0.19\tablenotemark{d}&      2.32&      3.31&      
3.09 \\
N/O &      0.52&      0.58&      0.44&      0.66&      0.21&      0.23&      
0.13&      0.11 \\
Ne/O&      0.21&      0.09&      0.07&      0.40&      6.76&      0.22&      
0.16&      0.15 \\
S/O ($\times 10^1$)&      0.05&      0.20&      0.18&      0.19&      0.09&      
0.07&      0.29&      0.28 \\
Cl/O ($\times 10^3$)&      0.29&      0.38&      0.40&      0.33&      \nodata&  
    0.29&      0.43&      0.41 \\
Ar/O ($\times 10^2$)  &      0.38&      0.77&      0.66\tablenotemark{d}&      
0.43\tablenotemark{d}&      0.03\tablenotemark{d}&      0.35&      0.45&      
0.59 \\

\enddata

\tablenotetext{a} {Uncertainties in elemental abundances are generally as 
follows. S/O: $\pm$30\%; Cl/O: $\pm$50\%; Ar/O: $\pm$75\%.}
\tablenotetext{b}{Grevesse et al. (1996)}
\tablenotetext{c}{Esteban et al. (1998), Table 19, gas + dust}
\tablenotetext{d}{Ar=Ar$^{+2}$}

\end{deluxetable}
\begin{deluxetable}{lcccccccc}
\tabletypesize{\scriptsize}
\setlength{\tabcolsep}{0.03in}
\tablecolumns{9}
\tablewidth{0in}
\tablenum{3E}
\tablecaption{Elemental Abundances\tablenotemark{a}}
\tablehead{
\colhead{Element} & 
\colhead{M1-54} &
\colhead{M1-57} &
\colhead{M2-10} &
\colhead{M3-4} &
\colhead{M3-6} &
\colhead{M3-15}   &
\colhead{Sun\tablenotemark{b}} &
\colhead{Orion\tablenotemark{c}}
}
\startdata

He/H&      0.15&      0.13&      0.13&      0.15&      0.12&      0.13&      
0.10&      0.10 \\
O/H ($\times 10^4$)&      5.91&      6.40&      6.05&      5.16&      5.59&      
6.82&      7.41&      5.25 \\
N/H ($\times 10^4$)&      5.70&      4.39&      3.01&      1.93&      0.58&      
3.38&      0.93&      0.60 \\
Ne/H ($\times 10^4$)&      2.41&      1.47&      1.52&      1.76&      1.36&     
 1.60&      1.20&      0.78 \\
S/H ($\times 10^5$)&      1.11&     0.74&      0.97&      0.16&      0.71&      
0.91&      2.14&      1.48 \\
Cl/H ($\times 10^7$)&      1.83&     2.47&      2.74&      0.84&      1.18&      
2.43&      3.16&      2.14 \\
Ar/H ($\times 10^6$)&      2.51\tablenotemark{d}&4.68       &      
3.45\tablenotemark{d}&      1.43\tablenotemark{d}&      3.06&      3.21&      
3.31&      3.09 \\
N/O &      0.96&      0.69&      0.50&      0.37&      0.10&      0.50&      
0.13&      0.11 \\
Ne/O&      0.41&      0.23&      0.25&      0.34&      0.24&      0.24&      
0.16&      0.15 \\
S/O ($\times 10^1$)&      0.19&      0.12&      0.16&      0.03&      0.13&      
0.13&      0.29&      0.28 \\
Cl/O ($\times 10^3$)&      0.31&      0.39&      0.45&      0.16&      0.21&     
 0.36&      0.43&      0.41 \\
Ar/O ($\times 10^2$)&      0.42\tablenotemark{d}&0.73       &      
0.57\tablenotemark{d}&      0.28\tablenotemark{d}&      0.55&      0.47&      
0.45&      0.59 \\

\enddata

\tablenotetext{a} {Uncertainties in elemental abundances are generally as 
follows. S/O: $\pm$30\%; Cl/O: $\pm$50\%; Ar/O: $\pm$75\%.}
\tablenotetext{b}{Grevesse et al. (1996)}
\tablenotetext{c}{Esteban et al. (1998), Table 19, gas + dust}
\tablenotetext{d}{Ar=Ar$^{+2}$}

\end{deluxetable}
\begin{deluxetable}{lcccccccc}
\tabletypesize{\scriptsize}
\setlength{\tabcolsep}{0.03in}
\tablecolumns{9}
\tablewidth{0in}
\tablenum{3F}
\tablecaption{Elemental Abundances\tablenotemark{a}}
\tablehead{
\colhead{Element} & 
\colhead{NGC 2792} &
\colhead{NGC 2867} &
\colhead{NGC 3195} &
\colhead{NGC 3211} &
\colhead{NGC 3242A} &
\colhead{NGC 3242B}   &
\colhead{Sun\tablenotemark{b}} &
\colhead{Orion\tablenotemark{c}}
}
\startdata

He/H&      0.11&      0.12&      0.13&      0.11&      0.11&      0.11&      
0.10&      0.10 \\
O/H ($\times 10^4$)&      7.65&      5.23&      6.88&      8.38&      4.37&      
3.83&      7.41&      5.25 \\
N/H ($\times 10^4$)&      1.39&      1.38&      3.09&      1.63&      0.51&      
0.47&      0.93&      0.60 \\
Ne/H ($\times 10^4$)&      1.26&      1.09&      2.77&      1.31&      0.90&     
 0.84&      1.20&      0.78 \\
S/H ($\times 10^5$)&      0.39&      0.33&      1.01&      0.44&      0.28&      
0.18&      2.14&      1.48 \\
Cl/H ($\times 10^7$)&      5.17&      1.47&      1.60&      4.52&      1.02&     
 1.10&      3.16&      2.14 \\
Ar/H ($\times 10^6$)&      6.42&      1.91&      2.74\tablenotemark{d}&      
6.33&      1.50&      1.50&      3.31&      3.09 \\
N/O&      0.18&      0.26&      0.45&      0.20&      0.12&      0.12&      
0.13&      0.11 \\ 
Ne/O&      0.16&      0.21&      0.40&      0.16&      0.21&      0.22&      
0.16&      0.15 \\
S/O ($\times 10^1$)&      0.05&      0.06&      0.15&      0.05&      0.06&      
0.05&      0.29&      0.28 \\
Cl/O ($\times 10^3$)&      0.68&      0.28&      0.23&      0.54&      0.23&     
 0.29&      0.43&      0.41 \\
Ar/O ($\times 10^2$)  &      0.84&      0.37&      0.40\tablenotemark{d}&      
0.76&      0.34&      0.39&      0.45&      0.59 \\

\enddata

\tablenotetext{a} {Uncertainties in elemental abundances are generally as 
follows. S/O: $\pm$30\%; Cl/O: $\pm$50\%; Ar/O: $\pm$75\%.}
\tablenotetext{b}{Grevesse et al. (1996)}
\tablenotetext{c}{Esteban et al. (1998), Table 19, gas + dust}
\tablenotetext{d}{Ar=Ar$^{+2}$}

\end{deluxetable}
\begin{deluxetable}{lcccccccc}
\tabletypesize{\scriptsize}
\setlength{\tabcolsep}{0.03in}
\tablecolumns{9}
\tablewidth{0in}
\tablenum{3G}
\tablecaption{Elemental Abundances\tablenotemark{a}}
\tablehead{
\colhead{Element} & 
\colhead{NGC 5307} &
\colhead{NGC 6309} &
\colhead{NGC 6439} &
\colhead{NGC 6563} &
\colhead{NGC 6565} &
\colhead{NGC 6629}   &
\colhead{Sun\tablenotemark{b}} &
\colhead{Orion\tablenotemark{c}}
}
\startdata

He/H&      0.10&      0.12&      0.14&      0.12&      0.12&      0.11&      
0.10&      0.10 \\
O/H ($\times 10^4$)&      3.85&      6.62&      6.45&      6.37&      7.25&      
4.50&      7.41&      5.25 \\
N/H ($\times 10^4$)&      0.51&      1.66&      4.59&      2.01&      2.96&      
0.62&      0.93&      0.60 \\
Ne/H ($\times 10^4$)&      0.90&      1.59&      1.88&      2.13&      2.41&     
 0.92&      1.20&      0.78 \\
S/H ($\times 10^5$)&      0.18&      0.50&      0.86&      0.33&      0.97&      
0.33&      2.14&      1.48 \\
Cl/H ($\times 10^7$)&      0.68&      3.42&      2.63&      0.30&      1.87&     
 0.95&      3.16&      2.14 \\
Ar/H ($\times 10^6$)&      1.36&      5.69&      4.75&      
1.89\tablenotemark{d}&      2.37\tablenotemark{d}&      1.88&      3.31&      
3.09 \\
N/O &      0.13&      0.25&      0.71&      0.32&      0.41&      0.14&      
0.13&      0.11 \\
Ne/O&      0.23&      0.24&      0.29&      0.33&      0.33&      0.20&      
0.16&      0.15 \\
S/O ($\times 10^1$)&      0.05&      0.08&      0.13&      0.05&      0.13&      
0.07&      0.29&      0.28 \\
Cl/O ($\times 10^3$)&      0.18&      0.52&      0.41&      0.05&      0.26&     
 0.21&      0.43&      0.41 \\
Ar/O ($\times 10^2$) &      0.35&      0.86&      0.74&      
0.30\tablenotemark{d}&      0.33\tablenotemark{d}&      0.42&      0.45&      
0.59 \\ 

\enddata

\tablenotetext{a} {Uncertainties in elemental abundances are generally as 
follows. S/O: $\pm$30\%; Cl/O: $\pm$50\%; Ar/O: $\pm$75\%.}
\tablenotetext{b}{Grevesse et al. (1996)}
\tablenotetext{c}{Esteban et al. (1998), Table 19, gas + dust}
\tablenotetext{d}{Ar=Ar$^{+2}$}

\end{deluxetable}
\begin{deluxetable}{lcccccc}
\tabletypesize{\scriptsize}
\setlength{\tabcolsep}{0.03in}
\tablecolumns{7}
\tablewidth{0in}
\tablenum{3H}
\tablecaption{Elemental Abundances\tablenotemark{a}}
\tablehead{
\colhead{Element} & 
\colhead{PB6} &
\colhead{PC14} &
\colhead{Pe1-18} &
\colhead{Th2-A}   &
\colhead{Sun\tablenotemark{b}} &
\colhead{Orion\tablenotemark{c}}
}
\startdata

He/H&      0.17&      0.12&      0.15&      0.13&      0.10&      0.10 \\
O/H ($\times 10^4$)&      6.77&      7.88&      4.44&      6.23&      7.41&      
5.25 \\
N/H ($\times 10^4$)&      7.64&      1.40&      9.15&      3.04&      0.93&      
0.60 \\
Ne/H ($\times 10^4$)&      1.60&      2.16&      1.20&      1.95&      1.20&     
 0.78 \\
S/H ($\times 10^5$)&      0.22&      0.74&      1.04&      0.41&      2.14&      
1.48 \\
Cl/H ($\times 10^7$)&      1.81&      2.08&      1.28&      0.78&      3.16&     
 2.14 \\
Ar/H ($\times 10^6$)&      5.86&      3.01&      3.57&      3.25&      3.31&     
 3.09 \\
N/O &      1.13&      0.18&      2.06&      0.49&      0.13&      0.11 \\
Ne/O&      0.24&      0.27&      0.27&      0.31&      0.16&      0.15 \\
S/O ($\times 10^1$)&      0.03&      0.09&      0.23&      0.07&      0.29&      
0.28 \\
Cl/O ($\times 10^3$)&      0.27&      0.26&      0.29&      0.12&      0.43&     
 0.41 \\
Ar/O ($\times 10^2$) &      0.87&      0.38&      0.80&      0.52&      0.45&    
  0.59 \\ 

\enddata

\tablenotetext{a} {Uncertainties in elemental abundances are generally as 
follows. S/O: $\pm$30\%; Cl/O: $\pm$50\%; Ar/O: $\pm$75\%.}
\tablenotetext{b}{Grevesse et al. (1996)}
\tablenotetext{c}{Esteban et al. (1998), Table 19, gas + dust}

\end{deluxetable}

\begin{deluxetable}{lccccccc}
\tablecolumns{7}
\tablewidth{0in}
\tablenum{4}
\tablecaption{Comparison of Abundance Averages}
\tablehead{
\colhead{Ratio} & 
\colhead{This Paper} &
\colhead{Type I} &
\colhead{Paper I} &
\colhead{KB\tablenotemark{1}} &
\colhead{AK\tablenotemark{2}} &
\colhead{Sun\tablenotemark{3}} &
\colhead{Orion\tablenotemark{4}}\\
\colhead{ } & 
\colhead{(all PNe)} &
\colhead{ } &
\colhead{(all PNe)} &
\colhead{ } &
\colhead{ } &
\colhead{ } &
\colhead{ }
}
\startdata
O/H (x 10$^4$) &5.4 $\pm$1.9& 5.9$\pm$.97& 5.5$\pm$1.5 & 4.8$\pm$2.0 & 4.4$\pm$.19  & 7.41 & 
5.25 \\
S/H (x 10$^5$) &0.56$\pm$.31& 0.73$\pm$.31& 0.69$\pm$.41 & 0.83$\pm$.82 & 1.1$\pm$.085  & 2.14 
& 1.48 \\
S/O (x 10$^1$) &0.11$\pm$.064&0.14$\pm$.62&0.13$\pm$.073&0.17$\pm$.14 & 
0.25$\pm$.022&0.29&0.28\\
Cl/H (x 10$^7$) &1.8$\pm$1.1&2.1$\pm$7.07& 1.9$\pm$1.1 & \nodata & 2.1$\pm$.18  & 3.16 & 2.14 
\\
Cl/O (x 10$^3$) &0.31$\pm$.12&0.36$\pm$.81&0.33$\pm$.15&\nodata&0.47$\pm$.04&0.43&0.41\\
Ar/H (x 10$^6$) &2.8$\pm$1.5&4.2$\pm$1.05& 2.8$\pm$1.2 & 2.5$\pm$2.5 & 2.9$\pm$.19 & 3.31 & 
3.09 \\
Ar/O (x 10$^2$) &0.51$\pm$.20&0.72$\pm$.16&0.51$\pm$.18&0.48$\pm$.48&0.69$\pm$.05&0.45&0.59
\enddata
\tablenotetext{1}{Average abundance ratios for a sample of planetary nebulae 
from Kingsburgh \& Barlow (1994)}
\tablenotetext{2}{Average abundance ratios for a sample of planetary nebulae 
from Aller \& Keyes (1987)}
\tablenotetext{3}{Grevesse et al. (1996)}
\tablenotetext{4}{Esteban et al. (1998), Table 19, gas + dust}
\end{deluxetable}

\clearpage

\begin{deluxetable}{lrrrrrrr}
\tablecolumns{8}
\tablewidth{0in}
\tablenum{5}
\tablecaption{Comparison of Abundances With Paper I\tablenotemark{a}}
\tablehead{
\colhead{Ratio} & 
\colhead{M1-50} &
\colhead{M1-54} &
\colhead{M1-57} &
\colhead{M3-15} &
\colhead{NGC 6309} &
\colhead{NGC 6439} &
\colhead{Pe1-18}
}
\startdata
He/H & 0.12/0.12 & 0.16/0.15 & 0.13/0.13 & 0.13/0.13 & 0.13/0.12 & 0.14/0.14 & 
0.15/0.15 \\
O/H($\times 10^4$) & 6.12/6.68 & 5.51/5.91 & 6.45/6.40 & 8.37/6.82 & 6.55/6.62 & 
6.19/6.49 & 4.68/4.44 \\
N/H($\times 10^4$) & 0.78/1.54 & 6.40/5.70 & 8.74/4.39 & 1.97/3.38 & 2.06/1.66 & 
5.56/4.59 & 10.45/9.15 \\
Ne/H($\times 10^4$) & 1.19/1.45 & 2.06/2.41 & 1.21/1.47 & 1.31/1.60 & 1.35/1.59 
& 1.62/1.88 & 0.74/1.20 \\
S/H($\times 10^5$) & 0.45/0.44 & 1.28/1.11 & 1.05/0.74 & 0.65/0.91 & 0.49/0.50 & 
1.31/0.86 & 0.94/1.04 \\
Cl/H($\times 10^7$) & 1.86/1.90 & 2.20/1.83 & 2.88/2.47 & 2.60/2.43 & 4.50/3.42 
& 2.90/2.63 & 2.39/1.28 \\
Ar/H($\times 10^6$) & 2.50/2.32 & 2.70/2.51 & 4.85/4.68 & 3.12/3.21 & 5.15/5.69 
& 2.06/4.75 & 3.07/3.57 \\
N/O & 0.13/0.23 & 1.16/0.96 & 1.35/0.69 & 0.23/0.50 & 0.32/0.25 & 0.90/0.71 & 
2.23/2.06 \\
Ne/O & 0.20/0.22 & 0.37/0.41 & 0.19/0.23 & 0.16/0.24 & 0.21/0.24 & 0.26/0.29 & 
0.16/0.27 \\
S/O($\times 10^1$) & 0.07/0.07 & 0.23/0.19 & 0.16/0.12 & 0.08/0.13 & 0.08/0.08 & 
0.21/0.13 & 0.20/0.23 \\
Cl/O($\times 10^3$) & 0.30/0.29 & 0.40/0.31 & 0.45/0.39 & 0.31/0.36 & 0.69/0.52 
& 0.47/0.41 & 0.51/0.29 \\
Ar/O($\times 10^2$) & 0.41/0.35 & 0.49/0.42 & 0.75/0.73 & 0.37/0.47 & 0.79/0.86 
& 0.33/0.74 & 0.65/0.80
\enddata

\tablenotetext{a}{ Paper I (Northern PNe) and this paper (southern PNe) have
seven objects in common. Numbers in 
each column are northern/southern.}
\end{deluxetable}

\clearpage

\begin{figure}
\figurenum{1} \plotone{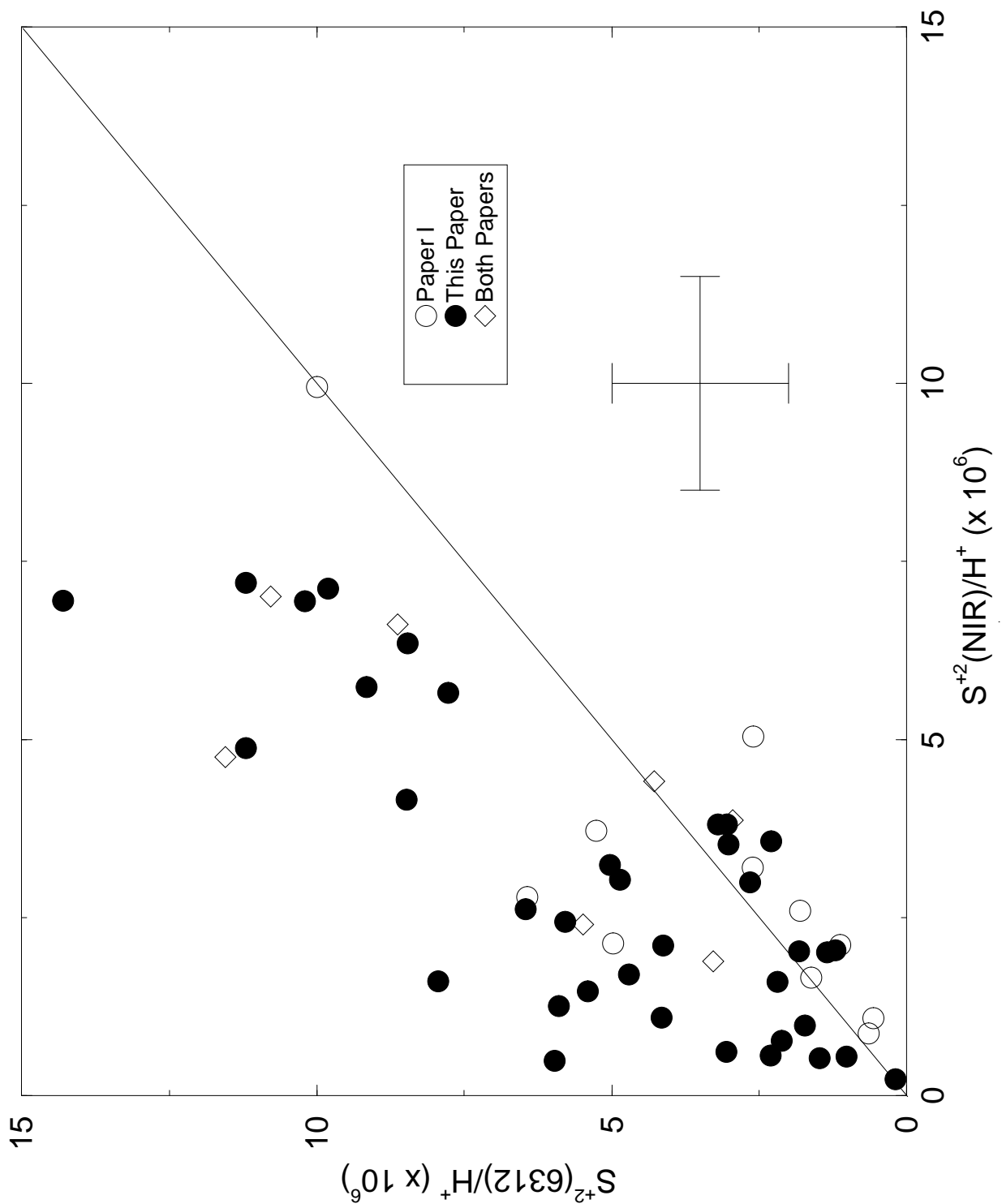}
\caption{Comparison of S$^{+2}$/H$^+$ for S abundances computed using
the 6312{\AA} emission line along with the [N~II] temperature
(ordinate) and the NIR emission lines along with the [S~III]
temperature (abscissa). Filled circles are objects in this paper, open
circles are objects in Paper~I, and diamonds are averages for objects which
are part of both papers. The solid line shows the track of a one-to-one
correspondence.}
\end{figure}

\begin{figure}
\figurenum{2}
\plotone{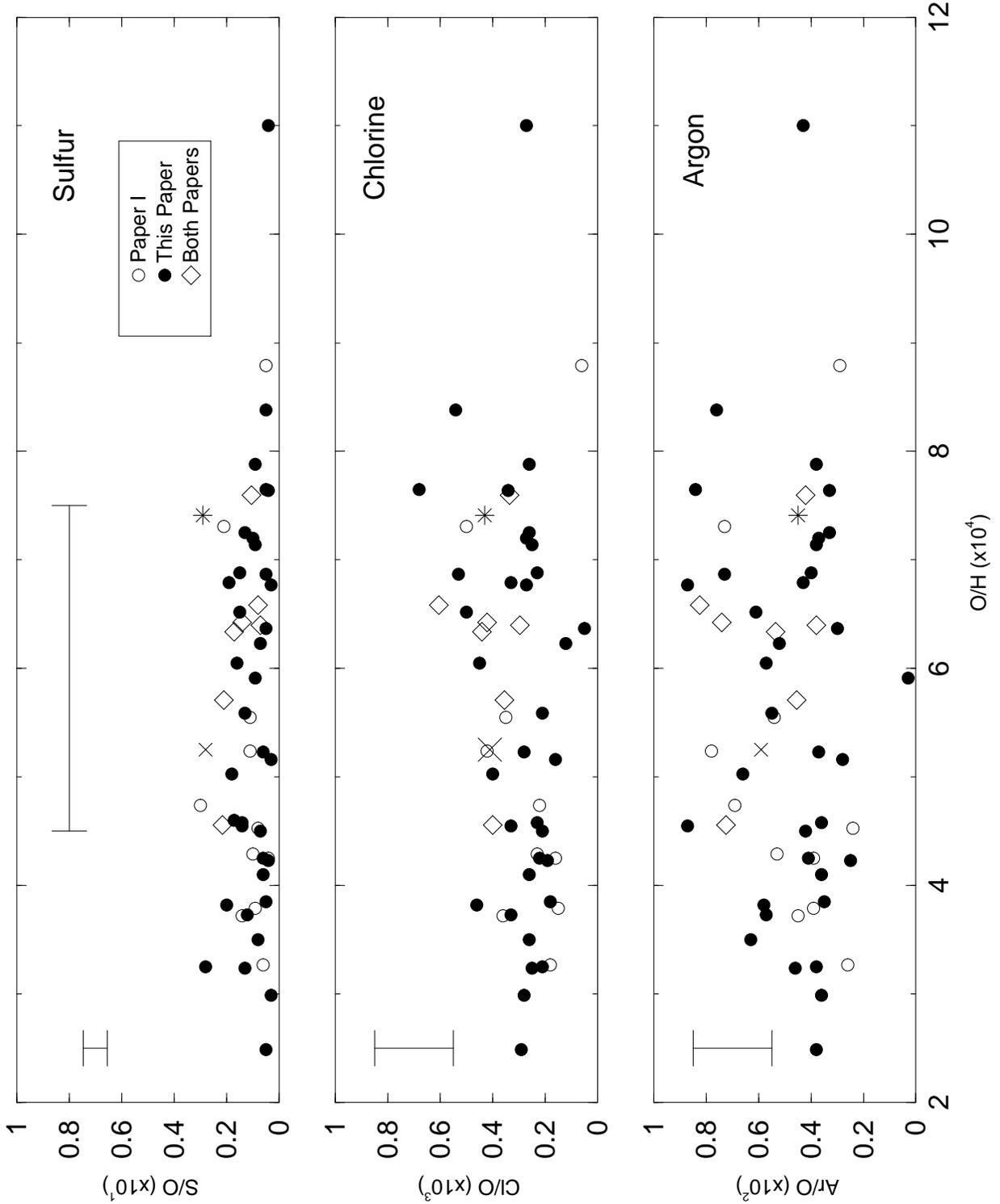}
\caption{Top: S/O (x 10$^1$) versus O/H (x 10$^4$), where filled
circles are ratios determined in this paper, open circles show
abundances determined in Paper~I, and diamonds show average values 
for objects included in both samples. The position of the Sun (Grevesse et
al. 1996) and the Orion Nebula (Esteban et al. 1998) are indicated
with a star and an X, respectively. Middle: Same as top but for Cl/O
(x 10$^3$). Bottom: Same as top but for Ar/O (x 10$^2$). Ordinate
uncertainties are shown with error bars in each panel, while the
horizontal error bar in the top panel shows the O/H uncertainty for
all three panels.}
\end{figure}

\end{document}